\documentclass[prl,english,twocolumn,superscriptaddress]{revtex4-2}
\usepackage[T1]{fontenc}
\usepackage{xcolor}
\usepackage{amssymb}
\usepackage[normalem]{ulem}
\usepackage{graphicx}

\usepackage{amsmath}
\usepackage{graphicx}
\usepackage{color}
\usepackage{bm}
\graphicspath{{IMAGES/}}
\usepackage{multirow}

\begin{document}
	
	\title{Anomalous acoustic plasmons in two-dimensional over-tilted Dirac bands}
	
	\author{Chang-Xu Yan}
	\thanks{These authors contributed equally to this work.}
	\affiliation{Department of Physics and Center for Computational Sciences, Sichuan Normal University, Chengdu, Sichuan 610066, China}
	
	\author{Furu Zhang}
	\thanks{These authors contributed equally to this work.}
	\affiliation{School of Science, Beijing Forestry University, Beijing 100083, China}
	
	\author{Chao-Yang Tan}
	\thanks{These authors contributed equally to this work.}
	\affiliation{Department of Physics and Center
		for Computational Sciences, Sichuan Normal University, Chengdu, Sichuan
		610066, China}
	
	\author{Hao-Ran Chang}
	\email{Corresponding author: hrchang@mail.ustc.edu.cn}
	\affiliation{Department of Physics and Center
		for Computational Sciences, Sichuan Normal University, Chengdu, Sichuan
		610066, China}
	
	\author{Jianhui Zhou}
	\email{Corresponding author: jhzhou@hmfl.ac.cn}
	\affiliation{Anhui Provincial Key Laboratory of Low-Energy Quantum Materials and Devices, High Magnetic Field Laboratory, HFIPS, Chinese Academy of Sciences, Hefei, Anhui 230031, China}
	
	\author{Yugui Yao}
	\affiliation{Key Laboratory of Advanced Optoelectronic Quantum Architecture and
		Measurement (MOE), School of Physics, Beijing Institute of Technology,
		Beijing 100081, China}

	\begin{abstract}
		The over-tilting of Dirac cones has led to various fascinating quantum phenomena. 
		Here we find that two anomalous acoustic plasmons (AAPs) are dictated by the distinct geometry of two-dimensional (2D) type-II Dirac cones, far beyond the conventional $\sqrt{q}$ plasmon. One AAP originates from the strong hybridization of two pockets with large velocity anisotropy at one Dirac point,
		whereas the other is attributed to the significant enhancement of the band correlation around the
		open Fermi surface. Remarkably, the plasmons exhibit valley-dependent chirality along the tilting
		direction due to the chiral electron dispersion. Meanwhile, we discuss the tunability of plasmon
		dispersion and lifetime by tuning the gap and dielectric substrate. Our work provides a promising
		way to generate the novel plasmons in Dirac materials.
	\end{abstract}
	
	\maketitle
	
	\textit{Introduction.--}Plasmon, the elementary excitation of electron liquids due to long-range Coulomb interaction,
	is a key ingredient of the emergent field of plasmonics \citep{pines1999elementary,maier2010plasmonics} and spintronics \citep{Raghu2010PRL}. Recently, plasmons of Dirac fermions with linear dispersion have been extensively investigated in various contents, such as  graphene-like systems \citep{Wunsch2006njp,Hwang2007PRB,Polini2008PRB,Gangadharaiah2008PRL,Cheng2017PRL,Wang2019prl,CaoPRL2021,Mojarro2022PRB}, surface states of three-dimensional (3D) topological insulators \citep{Raghu2010PRL,DiPietro.2013,Juergens2014PRL,ou2014plasmonTI,Kogar2015PRL,Politano2015PRL,Jia2017prl,ZhangFR2017PRL,Shvonski.2019,WeiT2022PRB}, 3D Dirac/Weyl semimetals \citep{DasSarmaPRL2009Plasmon3d,PanfilovPRB2014plasmon,Zhou2015Plasmon,Hofmann2015PRB,Kharzeev2015prl,Gorbar.2017,Politano2018PRL,Long2018PRL,Sadhukhan2020prl,Jia2020NJP,Wang2021pra,Heidari2021PRB,Afanasiev2021PRB}, and topological nodal line semimetals \citep{Yan.2016,Rhim.2016,Xue2021PRL,Islam2021PRB,Shao2022ZrSiSe}. Interestingly, the time-reversal symmetry breaking due to the external
	magnetic fields or spontaneous magnetization plays a crucial role in the creation of chiral or nonreciprocal plasmons \citep{Kumar2016PRB,Song2016PNAS,Mahoney2017NC,Lin2020PRL,Schlomer2021PRB}, facilitating the chiral optical responses. However, the studies of generation and manipulation of novel plasmons with respect to the emergent Lorentz invariance
	in low-dimensional Dirac materials possessing high tunability of carrier density and/or background dielectric
	constant are rarely reported.

\begin{figure}[htbp]
		\begin{centering}
			\includegraphics[scale=0.55]{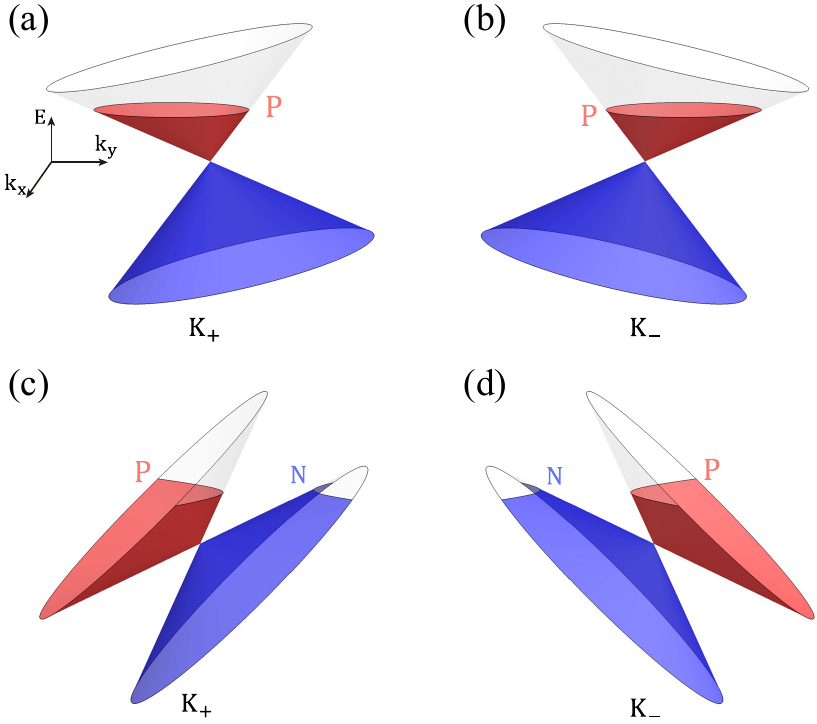}
		\end{centering}
		\caption{Energy dispersions and Fermi surfaces for (a,b) type-I $(0\le t<1)$ and (c,d) type-II $(t>1)$ Dirac semimetals ($\Delta=0$). A couple of Dirac cones in the $\mathrm{K}_{+}$ and $\mathrm{K}_{-}$ valleys tilt oppositely. The Fermi surfaces consist of (a,b) only one closed electron pocket (P) with $\lambda=+$ in type-I Dirac semimetals but (c,d) two open pockets---an electron pocket (P) with $\lambda=+$ and a hole pocket (N) with $\lambda=-$ in type-II Dirac semimetals.  
			\label{figure1}}
	\end{figure}

	It has been shown that, in contrast to type-I Dirac cones [Figs.\ref{figure1}(a,b)] with the point-like Fermi
	surface \citep{WanXGPRB2011,WengQAHE2015,Armitage2018rmp,Lv2021RMP}, the over-tilting of type-II Dirac cones [Figs.\ref{figure1}(c,d)] would break the Lorentz invariance and leads to mixed Dirac pockets at the Dirac nodes \citep{soluyanov2015t2}. Moreover, type-II Dirac/Weyl semimetals have been predicted to support a variety of novel electromagnetic responses \citep{Udagawa2016PRL,OBrienPRL2016,TchoumakovPRL2016,YuZMPRL2016,Halterman2018PRB} whose direct experimental evidences are still lacking. Surprisingly, recent works exhibited that the titling parameter can be largely tuned by delicate strains, triggering topological Lifshitz transitions between type-I and type-II Dirac semimetals \citep{zhang2017lifshitz} along with the dramatic changes of electronic states \citep{soluyanov2015t2,LiXPPRB2021}. Here we demonstrate that the alterable tilting of Dirac cones can be a new control knob of plasmons dictated by the unique geometry of type-II Fermi surface
	in a variety of two-dimensional (2D) Dirac materials, including the layered organic conductor $\alpha$-$(\mathrm{BEDT}$-$\mathrm{TTF})_{2}\mathrm{I}_{3}$ \citep{Bender1984mclc,Katayama2006jpsj,Hirata2016NC}, monolayer of
	transition metal dichalcogenides \citep{QianQSHEScience}, and 8-\emph{Pmmn} borophene \citep{ZhouXF2014PRL,Mannix2015Science,FengBJ2017PRL,Jalali2020PRB,LianC2020PRL} that possess multiple valleys in the Brillouin zone.

	In this work, we find the tilting of Dirac cones could give rise to three distinct tunable plasmons rather than a conventional
	intraband plasmon in 2D type-II Dirac materials. We reveal their origins in terms of band correlations within a new momentum-space two-component Dirac model. These plasmons propagating in the tilting direction are valley-dependent and chiral due to the chiral electron dispersion. In addition, increasing the background dielectric constant could shift
	the two higher frequency plasmons to lower frequency and change their lifetime, while increasing energy gap through electrical gating could merge them into one.
	
	\textit{Model.--}We begin with the effective Hamiltonian for a pair of 2D massive Dirac cones with tilting in the $y$-direction \citep{QianQSHEScience,Hirata2016NC,FengBJ2017PRL,tan2021PRB}
	\begin{eqnarray}
		H_{\chi}(\boldsymbol{k})&=\chi v_{t}k_{y}\tau_{0}+\chi v_{F}k_{x}\tau_{1}+v_{F}k_{y}\tau_{2}+\chi\Delta\tau_{3},\label{eq:Hamiltonian}
	\end{eqnarray}
	where $v_{F}$ is the Fermi velocity, $v_{t}$ denotes the band tilting, $\chi=\pm$ labels two valleys $\mathrm{K}_{\chi}$, $\Delta$ stands for the magnitude of mass which may associate with the strength of intrinsic spin-orbit coupling or the gap due to symmetry breaking of spatial inversion, and the reduced Planck's constant is set to be $\hbar=1$. Additionally, the unit matrix $\tau_{0}$ and Pauli matrix $\tau_{i}$ (with $i=1,2,3$) act upon the pseudo-spin space.
	The eigenvalues of Eq. \eqref{eq:Hamiltonian} are given as
	\begin{eqnarray}
		E_{\lambda}^{\chi}(\boldsymbol{k})&=\chi v_{t}k_{y}+\lambda\sqrt{v_{F}^{2}(k_{x}^{2}+k_{y}^{2})+\Delta^{2}},\label{eq:energydispersion}
	\end{eqnarray}
	where $\lambda=\pm$ are band indices. Tuning the tilting parameter $t\equiv v_{t}/v_{F}$, the Dirac semimetal would undergo a phase transition from type-I $(0\le t<1)$ Dirac semimetal [Figs.\ref{figure1}(a,b)] to type-II $(t>1)$ Dirac semimetal [Figs.\ref{figure1}(c,d)]. 
	
	Remarkably, each valley ($\mathrm{K}_{+}$ or $\mathrm{K}_{-}$) in the type-II Dirac semimetal possesses two open Fermi surfaces consisting of electron pocket (P) and hole pocket (N) characterized by distinct Fermi velocity along the tilting direction \citep{soluyanov2015t2} [Figs.\ref{figure1}(c,d)], in contrast
	to one closed elliptic Fermi surface, which consists of only one Dirac pocket in the type-I Dirac semimetal [Figs.\ref{figure1}(a,b)]. The highly anisotropic band structures and very distinct Fermi surfaces
	in the type-II Dirac semimetal would impose significant impacts on the plasmons.

	\textit{Anomalous plasmons.--}In order to reveal the nature of plasmon excitations for the type-II Fermi surface, we
	propose a model for tilted Dirac cones with multiple pockets in momentum space similar to the real-space model describing the two-component plasmons made of spatially-separated conventional 2D electron gases \citep{DasSarma1981PRB}. Accordingly, the total dielectric tensor takes
	\begin{align}
		\varepsilon_{\rho\lambda}(\boldsymbol{q},\omega)&= \delta_{\rho\lambda}-V_{\rho\lambda}(\boldsymbol{q})\sum_{\chi=\pm}\Pi_{\lambda}^{\chi}(\boldsymbol{q},\omega),
	\end{align}
	where $\rho$ and $\lambda$ are band indices. In this work, we focus on the electron correlations within each single Dirac point $\mathrm{K}_{\chi}$ due to the large separation between $\mathrm{K}_{+}$ and $\mathrm{K}_{-}$ in momentum space. For plasmons in multi-component system, the Coulomb interaction between carriers in different bands
	is taken to be $V_{\rho\lambda}(\boldsymbol{q})=V_{q}$, where $V_{q}=2\pi e^{2}/\kappa q$ is the Fourier transform of 2D
	Coulomb interaction with $\kappa$ the effective dielectric constant and $q=|\boldsymbol{q}|=\sqrt{q_{x}^{2}+q_{y}^{2}}$. The polarization function related to the band $\lambda$ around Dirac node $\mathrm{K}_{\chi}$ reads
	\begin{align}
		\Pi_{\lambda}^{\chi}(\boldsymbol{q},\omega)= & \sum_{\lambda^{\prime}=\pm\lambda}\int\frac{d^{2}\boldsymbol{k}}{(2\pi)^{2}}\mathcal{F}_{\lambda\lambda^{\prime}}^{\chi}(\boldsymbol{k},\boldsymbol{k}^{\prime})\nonumber \\
		& \times\frac{n_{F}[E_{\lambda}^{\chi}(\boldsymbol{k})]-n_{F}[E_{\lambda^{\prime}}^{\chi}(\boldsymbol{k}^{\prime})]}{\omega+E_{\lambda}^{\chi}(\boldsymbol{k})-E_{\lambda^{\prime}}^{\chi}(\boldsymbol{k}^{\prime})+i\eta},\label{eq:PolFun}
	\end{align}
	where 
	\begin{align} 
		&\mathcal{F}_{\lambda\lambda^{\prime}}^{\chi}\left(\boldsymbol{k},\boldsymbol{k}^{\prime}\right)=\left|\left\langle \chi,\lambda,\boldsymbol{k}|\chi,\lambda^{\prime},\boldsymbol{k}^{\prime}\right\rangle \right|^{2}
		\nonumber\\
		&=\frac{1}{2}\left\{1+\lambda\lambda^{\prime}\frac{v_{F}^{2}\boldsymbol{k}\cdot\boldsymbol{k}^{\prime} }{\sqrt{\left[(v_{F}\boldsymbol{k})^2+\Delta^2\right]\left[(v_{F}\boldsymbol{k}^{\prime})^{2}+\Delta^2\right]}}\right\}
	\end{align}
	represents the overlap between wave functions with $\left|\chi,\lambda,\boldsymbol{k}\right\rangle $
	being the periodic part of Bloch wave function \citep{ZhangFR2017PRL} and $\boldsymbol{k}^{\prime}=\boldsymbol{k}+\boldsymbol{q}$, $n_{F}(x)=\left[\exp\left[\beta(x-\mu)\right]+1\right]^{-1}$
	is the Fermi distribution function with $\beta=1/(k_{B}T)$ and $\mu$
	the chemical potential. It is emphasized that $\Pi_{\lambda}^{\chi}(\boldsymbol{q},\omega)$
	is contributed not only from the intraband process ($\lambda^{\prime}=+\lambda$)
	but also from the interband process ($\lambda^{\prime}=-\lambda$)
	and that the total dielectric tensor $\varepsilon_{\rho\lambda}(\boldsymbol{q},\omega)$
	receives contributions from both $\mathrm{K}_{+}$ and $\mathrm{K}_{-}$
	valleys. The condition for existence of collective modes is determined
	by the pole of the inverse of dielectric matrix $\varepsilon_{\rho\lambda}(\boldsymbol{q},\omega)$,
	leading to an effective dielectric function
	\begin{eqnarray}
		\varepsilon(\boldsymbol{q},\omega)\equiv 1-V_{q}\sum_{\chi=\pm}\sum_{\lambda=\pm}\Pi_{\lambda}^{\chi}(\boldsymbol{q},\omega).
	\end{eqnarray}
	The plasmon excitations can be equivalently identified by peaks in the electron energy-loss function (EELF), defined as the negative imaginary part of the inverse dielectric function, $\mathrm{Loss}(\boldsymbol{q},\omega)=\mathrm{Im}\left[-1/\text{\ensuremath{\varepsilon}}(\boldsymbol{q},\omega)\right],$
	which can be probed in various spectroscopy techniques, such as the electron energy-loss spectroscopy \citep{ibach.1982}, scattering-type scanning near-field optical microscopy \citep{Hesp2021NP}, and Fourier transform infrared spectroscopy \citep{grigorenko2012NPhot}. 
	
	\begin{figure}[htbp]
		\begin{centering}
			\includegraphics[scale=0.43]{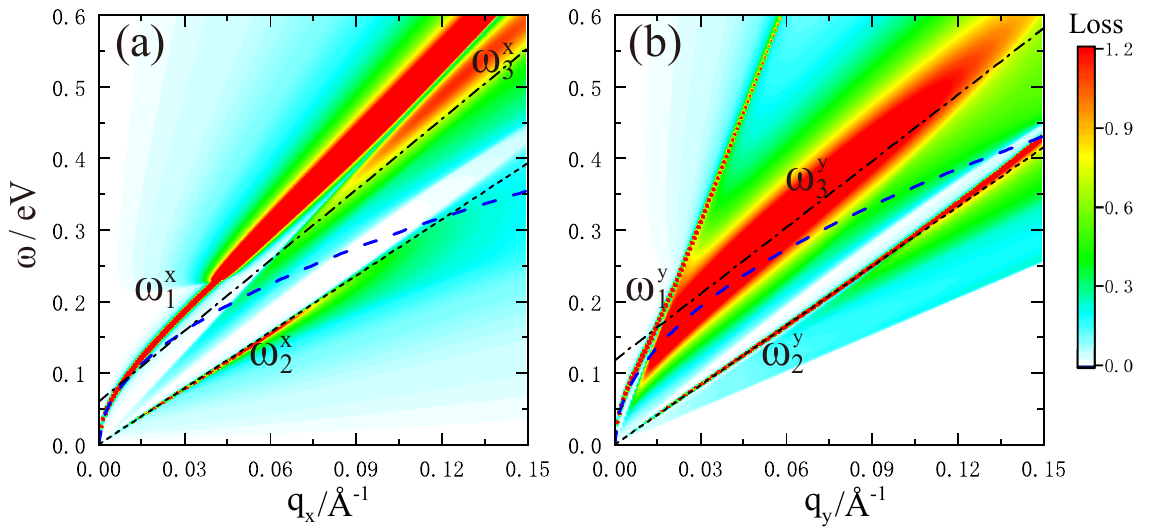}
			\par\end{centering}
		\caption{Plasmon dispersions from the EELF $\mathrm{Loss}(\boldsymbol{q},\omega)$. The approximate plasmon dispersions in Eqs.(\ref{omega2x})-(\ref{omega3y}) and the conventional $\sqrt{q}$ law \citep{SMs} are shown
			as dashed lines for comparison. The wave vector $\boldsymbol{q}=(q_{x},q_{y})$ is either (a) perpendicular to or (b) parallel to the tilting direction. Other parameters are taken as: Fermi velocity $v_{F}=0.65\times10^{6}\,\mathrm{m/s}$, tilting parameter $t=1.4$, chemical potential $\mu=0.25\,\mathrm{eV}$, dielectric constant $\kappa=15$, and cutoff of wave vector  $\Lambda=\pi/8\,\mathrm{\r{A}}^{-1}$ \citep{MuPlasmon}.
			\label{fig:qdependence}}
	\end{figure}

	Without loss of any generality, we at first focus on the gapless Dirac semimetals and discuss the influence of energy gap later. When the wave vector is perpendicular to [Figs.\ref{fig:qdependence}(a)] or parallel to [Fig.\ref{fig:qdependence}(b)]
	the tilting direction, there are three distinct plasmons \citep{TBPlasmon}. First, the plasmon $\omega_{1}^{x/y}(q)$ is a conventional mode that obeys the $\sqrt{q}$ law in the small wave vector regime, which is a common feature of plasmons in 2D metallic systems \citep{AndoRMP2deg}. Second, the first unusual plasmon $\omega_{2}^{x/y}(q)$ is an acoustic mode linear in $q$ which ubiquitously appears in the type-II Dirac semimetal. More interestingly, the intermediate frequency plasmon
	$\omega_{3}^{x/y}(q)$ near the $\sqrt{q}$ plasmon is quite unconventional, which has yet been discussed before in the conventional electron systems \citep{giuliani2005qtel} and Dirac materials \citep{Sadhukhan2020prl,Xue2021PRL}. Note that the plasmon with $q$-linear dispersion $\omega_{3}^{x/y}(q)$ usually hides as a shoulder in the region of intraband single particle excitation (SPE), hence dubbed the hidden plasmon in this work. 
	
\begin{figure}[htbp]
		\begin{centering}
			\includegraphics[scale=0.36]{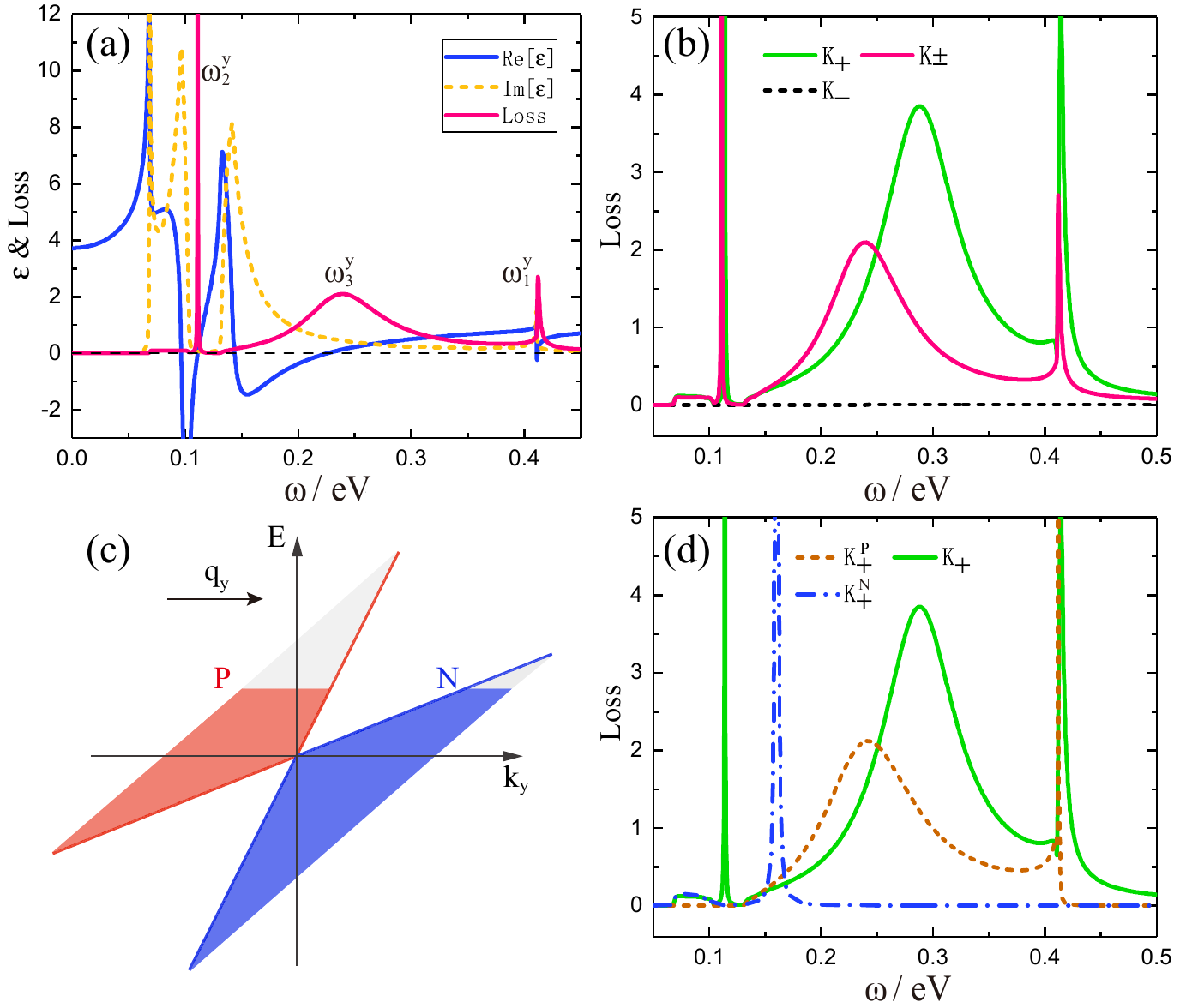}
			\par\end{centering}
		\caption{(a) Three plasmons are confirmed by the EELF and dielectric functions.
			(b) Three plasmons are contributed dominantly by the $\mathrm{K_{+}}$
			valley. (c) Schematic diagram of electron excitations in electron pocket (P) and hole pocket (N) at the $\mathrm{K_{+}}$
			valley. (d) Origin of plasmon peaks in terms of band correlations.
			The wave vector is $\boldsymbol{q}=(0,q_{y})$ with $q_{y}=0.04\:\text{\AA}^{-1}$.
			\label{fig:chiralplasmon}}
	\end{figure}

	\textit{Origin of plasmons.--}To unveil the origins of these plasmons in the type-II Dirac semimetals, we perform analysis based on the band correlations within the momentum-space two-component Dirac fermion model. Firstly, we consider the three
	plasmons $\omega_{1,2,3}^{y}(q)$ in Fig.\ref{fig:qdependence}(b). By means of decomposing the dielectric function into different isolated components \citep{ZhangFR2017PRL,Xue2021PRL,LiPRB2023}, numerical results of the EELF clearly reveal the three plasmon peaks in the type-II Dirac semimetals are dominated by the band correlation within the $\mathrm{K}_{+}$ valley, while the $\mathrm{K}_{-}$ valley does not contribute measurable signature [Figs.\ref{fig:chiralplasmon}(a,b)]. Thus, the plasmon modes exhibit valley-dependent chirality, which can be well understood within the chiral energy dispersion. Since along the tilting direction, the carriers that bear the type-II Dirac semimetal are essentially the 1D chiral Dirac fermions similar to
	the edge state in the quantum anomalous Hall system \citep{Yu2010Science,chang2013QAHEScience} and the corresponding energy dispersion is odd in $k_{y}$, that is, $E_{\lambda}^{\chi}(0,k_{y})=v_{F}(\chi tk_{y}+\lambda|k_{y}|)$
	[see the red and blue boundaries in Fig.\ref{fig:chiralplasmon}(c) for $\chi=+$ and $t>1$]. The resulting band correlations with $\boldsymbol{q}=(0,q_{y})$ can be achieved for supporting plasmon excitations in a given valley (for example $\mathrm{K}_{+}$), but not in the other valley with opposite tilting $\mathrm{K}_{-}.$ On the contrary, when the wave vector $\boldsymbol{q}=(q_{x},0)$ is perpendicular to the tilting direction, the resulting three plasmons
	$\omega_{1,2,3}^{x}(q)$ in the type-II Dirac semimetals [Fig.\ref{fig:qdependence}(a)] are contributed equally by two valleys with opposite tilting, independent of valley-selection, see the Supplemental Materials (SM) \citep{SMs}.
	Hence, we could focus on the contribution from the $\mathrm{K}_{+}$ valley hereafter.

	The conventional plasmons $\omega_{1}^{x}(q)$ and $\omega_{1}^{y}(q)$ are dominated by the intraband correlations within electron pocket due to
	the difference between electron pocket and hole pocket in size and velocity, as shown in the SM \citep{SMs} and Figs.\ref{fig:chiralplasmon}(c,d). Next we would examine the two anomalous acoustic plasmons (AAPs), $\omega_{2}^{x/y}(q)$ and $\omega_{3}^{x/y}(q)$. The lower-frequency AAPs $\omega_{2}^{x}(q)$ and $\omega_{2}^{y}(q)$ linear in $q$, similar to the gapless Pines' demon \citep{pines1956cjp}, originate from the strong hybridization or out-of-phase oscillations \citep{Afanasiev2022PRB} between intraband correlations
	from two pockets with different velocities [see the Supplemental Materials \citep{SMs} and Figs.\ref{fig:chiralplasmon}(c,d)]. Furthermore, it is worthy to emphasize that the AAP $\omega_{2}^{x}(q)$ in Fig.\ref{fig:qdependence}(a) is absent in the type-I Dirac semimetal since the plasmons therein become degenerate, sharing the same velocity \citep{Nishine2011JPSJ}. The approximate dispersions of these lower-frequency AAPs are given as
	\begin{align}
		\omega_{2}^{x}(q)&\approx\sqrt{1-\frac{1}{t^2}\left(1+\frac{|\mu|}{t v_{F}\Lambda}\right)^2}v_{F}q,\label{omega2x}\\
		\omega_{2}^{y}(q)&\approx\left[t-\frac{1}{t}\left(1+\frac{|\mu|}{2tv_{F}\Lambda}\right)\right]v_{F}q,\label{omega2y}
	\end{align}
	with $\Lambda$ being the cutoff of wave vector, which indicates that the titling can be used to tune the velocity of plasmon excitations (see the Supplemental Materials \citep{SMs}).
	
	\begin{figure}[htbp]
		\begin{centering}
			\includegraphics[scale=0.44]{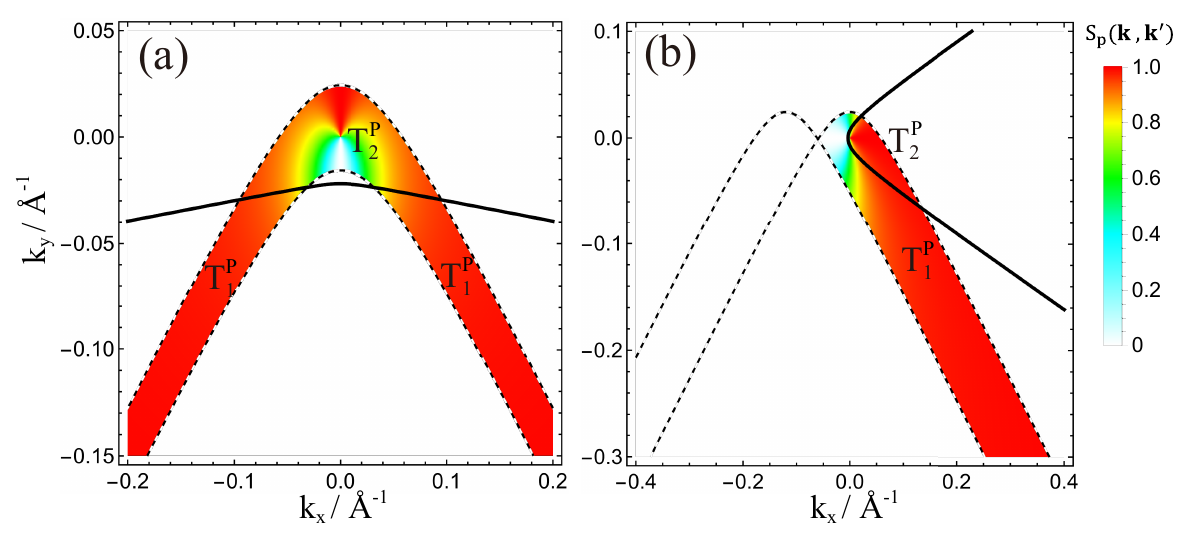}
			\par\end{centering}
		\caption{Origin of the hidden plasmons $\omega_{3}^{x/y}(\boldsymbol{q})$ living only in the intraband SPE region. $\mathcal{S}_{\mathrm{P}}(\boldsymbol{k},\boldsymbol{k}^{\prime})=\mathcal{F}_{++}^{+}(\boldsymbol{k},\boldsymbol{k}^{\prime})\ensuremath{\Theta\left[E_{+}^{+}(\boldsymbol{k}^{\prime})-\mu\right]}\ensuremath{\Theta\left[\mu-E_{+}^{+}(\boldsymbol{k})\right]}$ 
			is plotted  with a wave vector (a) $q_{y}=0.04~\text{\AA}^{-1}$ and (b) $q_{x}=0.12~\text{\AA}^{-1}$.
			The boundary between $T_1^P$ and $T_2^P$ regions is not predefined but is determined self-consistently by all electronic transitions, ultimately coinciding with $E_{+}^{+}(\boldsymbol{k}^{\prime})-E_{+}^{+}(\boldsymbol{k})=\omega_3(\boldsymbol{q})$.}
		\label{fig:omega3}
	\end{figure}

	The higher-frequency AAPs $\omega_{3}^{x}(q)$ and $\omega_{3}^{y}(q)$ are dominated by the intraband correlations of electrons in the electron pocket [see Figs.\ref{fig:chiralplasmon}(c,d)], and live only in the intraband SPE region [see Fig.\ref{fig:qdependence}]. This hardly occurs in conventional metals, because plasmons and intraband single-particle excitations cannot coexist with each other. However, it could be possible due to the unique overtilted band structure, which creates an extensive single-particle excitation (SPE) region characterized by an open boundary. This geometry provides a high density of electronic transitions that uniquely sustain a collective mode ($\omega_3$) within the SPE continuum. As shown in Fig. \ref{fig:omega3}, these transitions are partitioned into two subsets, $T_1^P$ and $T_2^P$, separated by a self-consistently determined energy threshold $E_{+}^{+}(\boldsymbol{k}^{\prime})-E_{+}^{+}(\boldsymbol{k})=\omega_3(\boldsymbol{q})$. 
	The $\omega_3$ mode emerges from the competition between the open-boundary region ($T_1^P$), which favor plasmon formation by a positive contribution to $\Pi_{\lambda}^{\chi}(\boldsymbol{q},\omega)$, and the closed-boundary region ($T_2^P$), which tend to suppress it by a negative contribution. Unlike standard systems with closed Fermi surfaces where $T_1^P$ is typically too weak to overcome the damping effect of $T_2^P$, the unique thin-strip geometry of this transition region drives the linear dispersion of $\omega_3$. Consequently, $\omega_3^x$ and $\omega_3^y$ modes serve as unambiguous evidence for the existence of 2D Type-II Dirac bands.

	The dispersions of these higher-frequency AAPs can be approximately captured by
	\begin{align}
		\hspace{-0.2cm}\omega_{3}^{x}(q)&\approx\frac{e^2\Lambda}{2\pi\kappa}+\sqrt{1-\frac{1}{t^2}\left(1-\frac{|\mu|}{t v_{F}\Lambda}\right)^2}v_{F}q,\label{omega3x}\\
		\hspace{-0.2cm}\omega_{3}^{y}(q)&\approx\sqrt{t^2-1}\frac{e^2\Lambda}{\pi\kappa}+\left[t-\frac{1}{t}\left(1-\frac{|\mu|}{2tv_{F}\Lambda}\right)\right]v_{F}q,\label{omega3y}
	\end{align}
	possessing cutoff-dependent gaps. In contrast to the gapless acoustic plasmon in the classical two-component model \citep{DasSarma1981PRB}, the gap of the AAP originates from the finite-range Coulomb interaction. The approximate dispersions of AAPs imply that the titling can be used to tune the velocity of plasmon excitations \citep{SMs}. Note that multiple plasmon excitations had been unveiled in 3D type-II Dirac cones \cite{Sadhukhan2020prl}. The different dimensionality and Coulomb screening cause distinct features between the triple plasmons here and those in 3D case \citep{SMs}.

	\begin{figure}[htbp]
		\begin{centering}
			\includegraphics[scale=0.43]{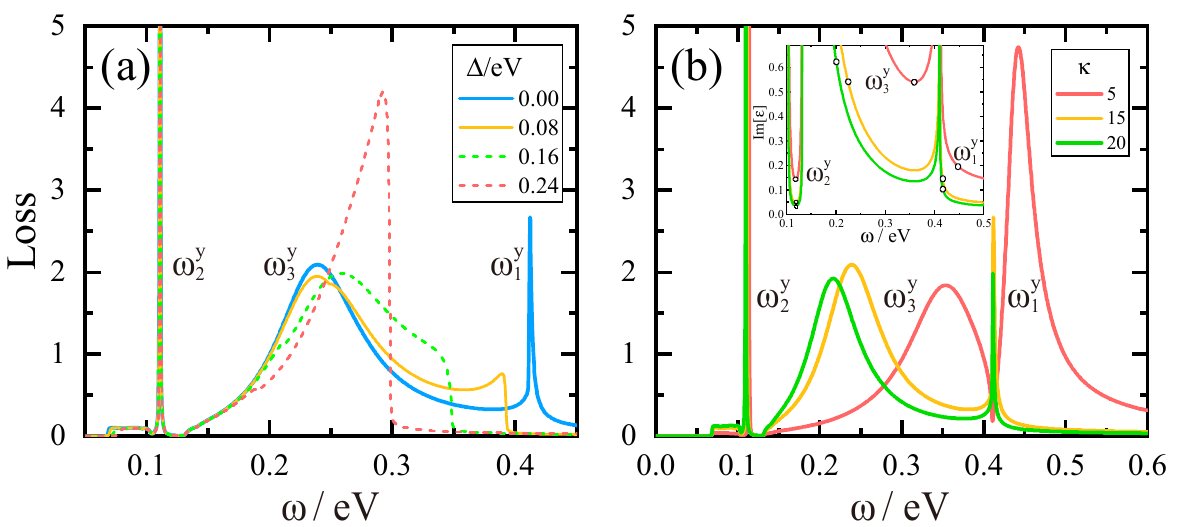}
			\par\end{centering}
		\caption{Impacts of (a) energy gap and (b) dielectric constant on plasmon modes
			$\omega_{1}^{y}(q)$, $\omega_{2}^{y}(q)$, and $\omega_{3}^{y}(q)$ with a wave vector $q_{y}=0.04\:\text{\AA}^{-1}$.
			Other parameters are set to be $\kappa=15$ in (a) and $\Delta=0$
			in (b). The imaginary part of $\text{\ensuremath{\varepsilon}}(\mathrm{\boldsymbol{q}},\omega)$
			is shown in the inset of (b), where the circles denote the corresponding
			plasmon frequencies for different dielectric constants.}
		\label{fig:gapdependence}
	\end{figure}
	
	\textit{Manipulations of plasmons.--}Let us turn to the impacts of energy gap and dielectric background on the plasmon modes in 2D type-II Dirac cones. With increasing the magnitude of energy gap (from $\Delta=0$ to $\Delta=0.24\:\mathrm{eV}$), the
	two higher frequency plasmons $\omega_{\text{1}}^{y}(q)$ and $\omega_{3}^{y}(q)$ would eventually merge into one strong plasmon [Fig.\ref{fig:gapdependence}(a)]. In addition, increasing the effective dielectric constant would continuously
	shift the peaks of the two higher frequency plasmons to lower frequency [Fig.\ref{fig:gapdependence}(b)], and change their lifetime [see the inset in Fig.\ref{fig:gapdependence}(b)]. Interestingly, the peak of the lowest plasmon $\omega_{\text{2}}^{y}(q)$ remains almost unchanged against the change of energy gap or dielectric background [Fig.\ref{fig:gapdependence}], but the lifetime of this plasmon is strengthened with increasing the dielectric constant [see the inset in Fig.\ref{fig:gapdependence}(b)]. In fact, the magnitude of energy gap can be effectively modified by the gate voltages \citep{QianQSHEScience,Wu2018Science}, and the dielectric constant can be changed with different substrates, allowing us to manipulate the plasmon dispersions. The highly tunable plasmons would enable promising applications of 2D Dirac materials in plasmonics.
	
	\textit{Conclusions.--}In summary, we demonstrated that the tunable over-tilting of 2D Dirac cones leads to two AAPs associated with the distinct geometry of Fermi surface in type-II Dirac semimetals. One of the AAPs results from the strong hybridization of two pockets at a common Dirac point. Notably, this over-tilting enhances the band correlation to activate a strong unusual plasmon $\omega_{3}^{x/y}(q)$ that could characterize type-II Dirac bands. Moreover, the unique chiral electron-hole excitations along the tilting direction yield valley-nonreciprocity of plasmon
	modes. Thus, we hope that the present predictions would inspire immediate experimental investigations of the novel plasmons in various Dirac materials and their plasmonic applications. 
	
	The authors thank J.W. Liu for useful discussions. This work was financially supported by the National Key R\&D Program of the MOST of China (Grant No. 2024YFA1611300 and No. 2020YFA0308800), the National Natural Science Foundation of China (Grant No. 11547200, No. 12574059 and No. 12174394), HFIPS Director's Fund (Grant No. BJPY2023B05), Anhui Provincial Major S\&T Project (s202305a12020005), the Basic Research Program of the Chinese Academy of Sciences Based on Major Scientific Infrastructures (Grant No. JZHKYPT-2021-08), the High Magnetic Field Laboratory of Anhui Province under Contract No. AHHM-FX-2020-02, the China Postdoctoral Science Foundation (Grant No. 2019M650583), and the Research Institute of Intelligent Manufacturing Industry Technology of Sichuan Arts and Science University. We thank the High Performance Computing Center at Sichuan Normal University.

			\begin{widetext}
				
				\makeatletter
				\renewcommand \thesection{S\@arabic\c@section}
				\renewcommand\thetable{S\@arabic\c@table}
				\renewcommand \thefigure{S\@arabic\c@figure}
				\renewcommand \theequation{S\@arabic\c@equation}
				\makeatother
			
			\begin{center}
				{\textbf{Supplemental Materials to ``Anomalous acoustic plasmons in two-dimensional over-tilted Dirac bands''}}
			\end{center}

			
			\section{S1. Impact of chemical potential on the plasmons}
			
			\begin{figure}[h]
				\includegraphics[scale=0.53]{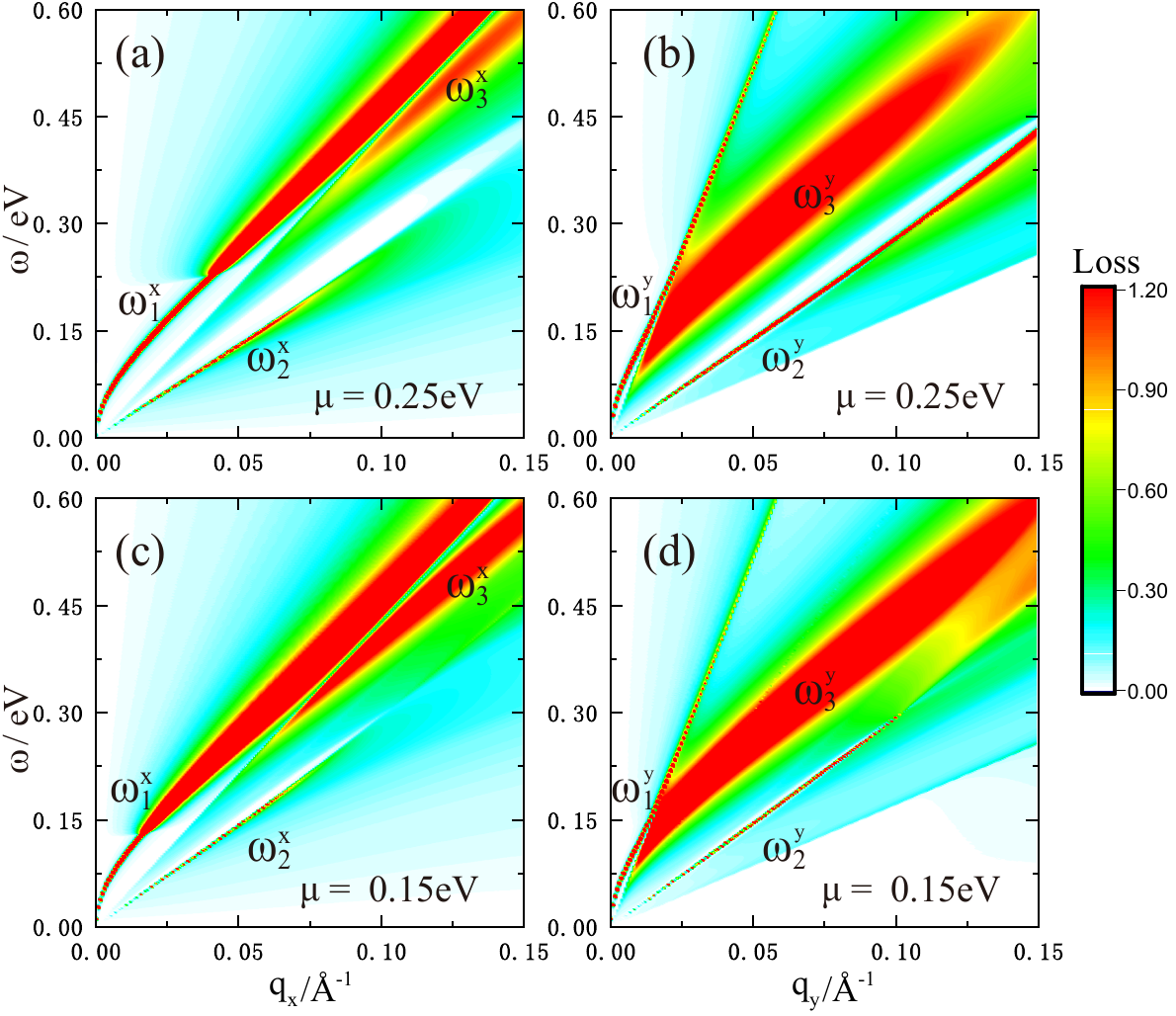}
				\caption{Impact of chemical potential on the plasmons in 2D type-II Dirac semimetal. The wave vector $\boldsymbol{q}$ is perpendicular to the tilting direction in (a) and (c), or parallel to the tilting direction in (b) and (d). Other parameters are taken to be the same as those in the main text.}
				\label{fig:mudependence}
			\end{figure}
			
			The impact of chemical potential on the plasmon modes in 2D type-II Dirac semimetal is shown in Fig. \ref{fig:mudependence}. It is obvious that the chemical potential does not substantially influence the qualitative properties of plasmon dispersions.

			\section{S2. Plasmon dispersions for 2D type-II Dirac bands in the tight-binding model}
			
			To confirm our results of plasmon dispersions in the $\emph{k}\cdot\emph{p}$ Hamiltonian for 2D type-II Dirac cone,
			we present a calculation based on the equivalent tight-binding Hamiltonian
			\begin{equation}
				H(\boldsymbol{k})=\frac{\upsilon_{t}}{a}\cos(k_{y}a)\tau_{0}+\frac{\upsilon_{F}}{a}\sin(k_{x}a)\tau_{1}+\frac{\upsilon_{F}}{a}\cos(k_{y}a)\tau_{2}+\Delta\tau_{3},
			\end{equation}
			whose energy dispersion reads
			\begin{equation}
				E_{\lambda}(\boldsymbol{k})=\frac{1}{a}\left\{ \upsilon_{t}\cos(k_{y}a)+\lambda\sqrt{\upsilon_{F}^{2}\left[\sin^{2}(k_{x}a)+\cos^{2}(k_{y}a)\right]+a^{2}\Delta^{2}}\right\} ,
			\end{equation}
			which possesses two band crossings at $\mathrm{K}_{\chi}=(0,\chi\frac{\pi}{2a})$
			with $a$ the lattice spacing. This tight-binding Hamiltonian reduces to the $\emph{k}\cdot\emph{p}$ Hamiltonian for 2D type-II Dirac cone in the main text.
			
			\begin{figure}[th]
				\includegraphics[scale=0.53]{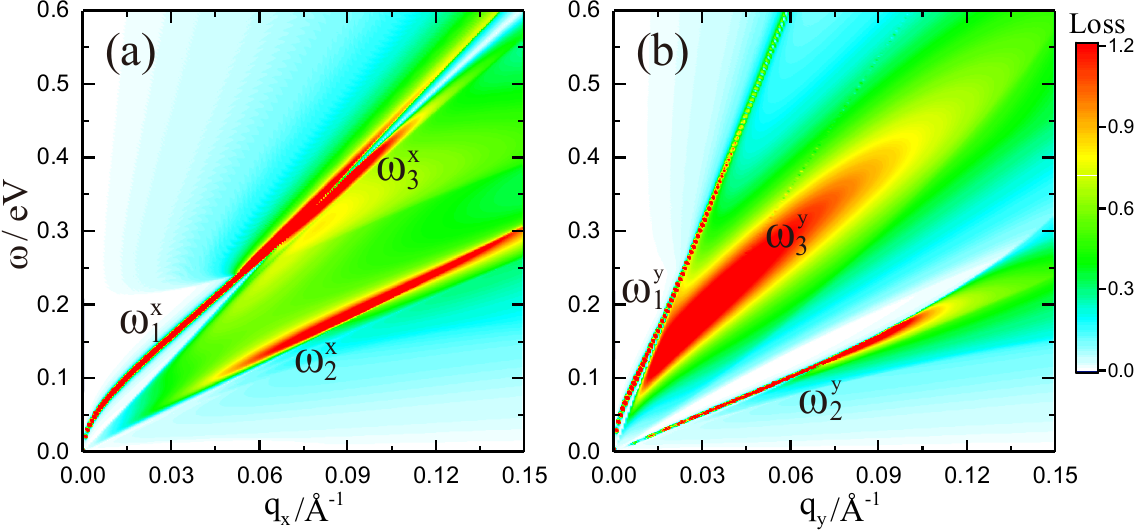}
				\caption{Plasmon dispersions for 2D type-II Dirac semimetal in the tight-binding model. The wave vector $\boldsymbol{q}$ is (a) perpendicular to the tilting direction, or (b) parallel to the tilting direction. The lattice spaceing $a=4\mathrm{\mathring{A}}$. Other parameters are taken to be the same as those in the main text. \label{fig:TB}}
			\end{figure}

			From the Matasubara Green's function in momentum space
			\begin{equation}
				\mathcal{G}(\boldsymbol{k},i\Omega_{n})=\frac{1}{2}\sum_{\lambda=\pm}\cfrac{1}{i\Omega_{n}+\mu-E_{\lambda}(\boldsymbol{k})}\left[\tau_{0}+\lambda\text{\ensuremath{\cfrac{\upsilon_{F}\sin(k_{x}a)\tau_{1}+\upsilon_{F}\cos(k_{y}a)\tau_{2}+a\Delta\tau_{3}}{\sqrt{\upsilon_{F}^{2}\left[\sin^{2}(k_{x}a)+\cos^{2}(k_{y}a)\right]+a^{2}\Delta^{2}}}}}\right],
			\end{equation}
			one can arrive at the corresponding polarization function
			\begin{align}
				\text{\ensuremath{\Pi}(\ensuremath{\boldsymbol{q}},\ensuremath{\omega})} & =\frac{1}{V}\sum_{\boldsymbol{k}}\frac{1}{\beta}\sum_{i\Omega_{n}}\mathrm{Tr}\left[\mathcal{G}(\boldsymbol{k},i\Omega_{n})\mathcal{G}(\boldsymbol{k}+\boldsymbol{q},i\Omega_{n}+\omega+i\eta)\right]=\sum_{\lambda,\lambda^{\prime}=\pm}\text{\ensuremath{\Pi}}_{\lambda\lambda^{\prime}}(\boldsymbol{q},\omega),
			\end{align}
			where
			\begin{align}
				\text{\ensuremath{\Pi}}_{\lambda\lambda^{\prime}}(\boldsymbol{q},\omega) & =\frac{1}{V}\sum_{\boldsymbol{k}}\mathcal{F}_{\lambda\lambda^{\prime}}(\boldsymbol{k},\boldsymbol{k}+\boldsymbol{q})\frac{n_{F}[E_{\lambda}(\boldsymbol{k})]-n_{F}[E_{\lambda^{\prime}}(\boldsymbol{k}+\boldsymbol{q})]}{\omega+E_{\lambda}(\boldsymbol{k})-E_{\lambda^{\prime}}(\boldsymbol{k}+\boldsymbol{q})+i\eta},
			\end{align}
			with
			\begin{align}
				&\mathcal{F}_{\lambda\lambda^{\prime}}(\boldsymbol{k},\boldsymbol{k}+\boldsymbol{q})
				\nonumber\\&=
				\frac{1}{2}\left\{ 1+\lambda\lambda^{\prime}\frac{\upsilon_{F}^{2}\left\{ \sin(k_{x}a)\sin\left[(k_{x}+q_{x})a\right]+\cos(k_{y}a)\cos\left[(k_{y}+q_{y})a\right]\right\} +a^{2}\Delta^{2}}{\sqrt{\upsilon_{F}^{2}\left[\sin^{2}(k_{x}a)+\cos^{2}(k_{y}a)\right]+a^{2}\Delta^{2}}\sqrt{\upsilon_{F}^{2}\left\{ \sin^{2}\left[(k_{x}+q_{x})a\right]+\cos^{2}\left[(k_{y}+q_{y})a\right]\right\} +a^{2}\Delta^{2}}}\right\} .
			\end{align}
			Replacing $\frac{1}{V}\sum_{\boldsymbol{k}}$ with $\int\frac{d^{2}\boldsymbol{k}}{(2\pi)^{2}}$,
			one then obtains the corresponding polarization function as
			\begin{align}
				\text{\ensuremath{\Pi}}_{\lambda\lambda^{\prime}}(\boldsymbol{q},\omega)= & \int\frac{d^{2}\boldsymbol{k}}{(2\pi)^{2}}\mathcal{F}_{\lambda\lambda^{\prime}}(\boldsymbol{k},\boldsymbol{k}+\boldsymbol{q})\frac{n_{F}[E_{\lambda}(\boldsymbol{k})]-n_{F}[E_{\lambda^{\prime}}(\boldsymbol{k}+\boldsymbol{q})]}{\omega+E_{\lambda}(\boldsymbol{k})-E_{\lambda^{\prime}}(\boldsymbol{k}+\boldsymbol{q})+i\eta}.
			\end{align}

			By finding the zeros of dielectric function
			\begin{align}
				\varepsilon(\boldsymbol{q},\omega)=1-V_{q}\text{\ensuremath{\Pi}(\ensuremath{\boldsymbol{q}},\ensuremath{\omega})}=1-V_{q}\sum_{\lambda,\lambda^{\prime}=\pm}\text{\ensuremath{\Pi}}_{\lambda\lambda^{\prime}}(\boldsymbol{q},\omega)=0,
			\end{align}
			one can obtain the plasmon dispersions in the tight-binding model
			for 2D type-II Dirac cone.

			The plasmon dispersions in the tight-binding model for 2D type-II Dirac semimetal ($\Delta=0$) are shown in Fig. \ref{fig:TB}. Compared with the plasmon dispersions in the $k\cdot p$ Hamiltonian for 2D type-II Dirac semimetal in Fig. \ref{fig:mudependence}(a,b) [see also Fig.2 in the main text], the corresponding tight-binding Hamiltonian for 2D type-II Dirac semimetal yields consistent plasmon dispersions, which indicates that the essential physics can be well captured by the $k\cdot p$ Hamiltonian in the main text.

			\section{S3. Origin of plasmons in the 2D type-II Dirac semimetal perpendicular to the tilting direction}
			
			The origin of plasmon in type-II Dirac semimetal are revealed in terms
			of band correlations when the wave vector $\boldsymbol{q}=(q_{x},0)$
			is perpendicular to the direction of tilting in Fig. \ref{fig:type-IIOrigin-qx1}.
			
			\begin{figure}[htbp]
				\includegraphics[scale=0.55]{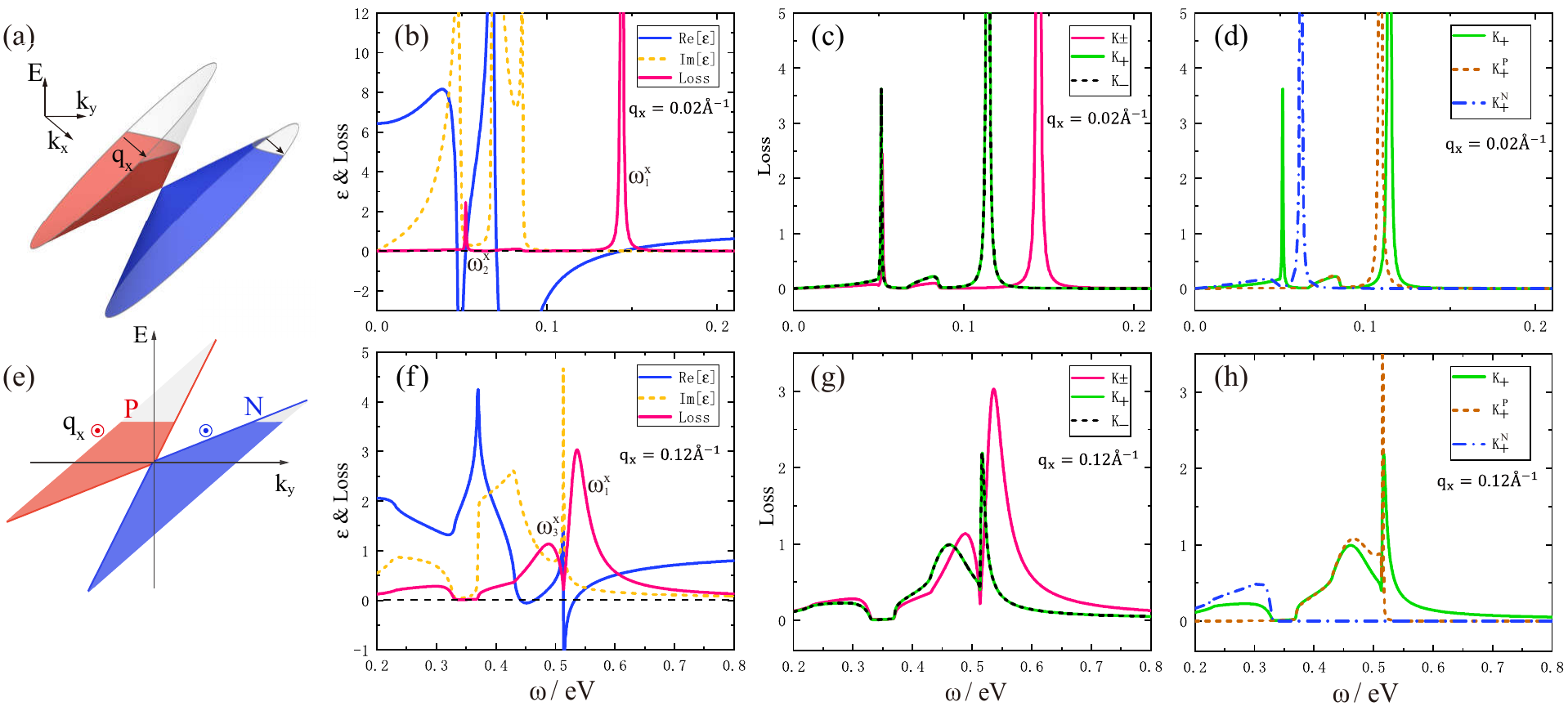}
				\caption{Origin of plasmons in 2D type-II Dirac semimetal in terms of band correlations. Panels (a) and (e) denote the corresponding energy dispersion. The wave vector $\boldsymbol{q}=(q_{x},0)$ is perpendicular to the direction of tilting with (b-d) $q_{x}=0.02\mathrm{\mathring{A}^{-1}}$ and (f-h) $q_{x}=0.12\mathrm{\mathring{A}^{-1}}$, and other parameters are taken to be the same as those in main text. \label{fig:type-IIOrigin-qx1}}
			\end{figure}

			These figures indicate that the essential features of plasmons can
			be well captured by band correlations at the valley around Dirac node
			$\mathrm{K}_{+}$ when the wave vector $\boldsymbol{q}=(q_{x},0)$
			is perpendicular to the direction of tilting, similar to that when
			the wave vector $\boldsymbol{q}=(0,q_{y})$ is parallel to the titling direction
			[see Fig. 3 in the main text]. Hereafter, we are allowed
			to restrict our physical analysis to the valley around Dirac node
			$\mathrm{K}_{+}$. After further dividing the bands at the valley
			around Dirac point $\mathrm{K}_{+}$ into $\mathrm{K}_{+}^{\mathrm{P}}$
			($\lambda=+$) and $\mathrm{K}_{+}^{\mathrm{N}}$ ($\lambda=-$),
			one could find that the plasmons $\omega_{1}^{x/y}(q)$ and $\omega_{3}^{x/y}(q)$
			are contributed by $\mathrm{K}_{+}^{\mathrm{P}}$, but the plasmon
			$\omega_{2}^{x/y}(q)$ is strongly related to the hybridization between
			two pockets $\mathrm{K}_{+}^{\mathrm{P}}$ ($\lambda=+$) and $\mathrm{K}_{+}^{\mathrm{N}}$
			($\lambda=-$), although it is contributed dominantly by the pocket
			$\mathrm{K}_{+}^{\mathrm{N}}$ [see Fig.\ref{fig:type-IIOrigin-qx1}(d)].

			\section{S4. Origin of plasmons in the 2D type-I Dirac semimetal}
			
			As a comparison, we provide the plasmon dispersion for 2D type-I Dirac
			semimetal, and reveal the origin in terms of the hybridization between
			between the $\mathrm{K}_{+}$ and $\mathrm{K}_{-}$ valleys in Fig. \ref{fig:type-IOrigin}.
			The plasmon dispersion for 2D type-I Dirac semimetal has been found
			previously \cite{JPSJNishine2011}, which provides a theoretical mechanism to realize
			the Pines' demon \cite{CJPPines1956}.

			\begin{figure}[htbp]
				\includegraphics[scale=0.65]{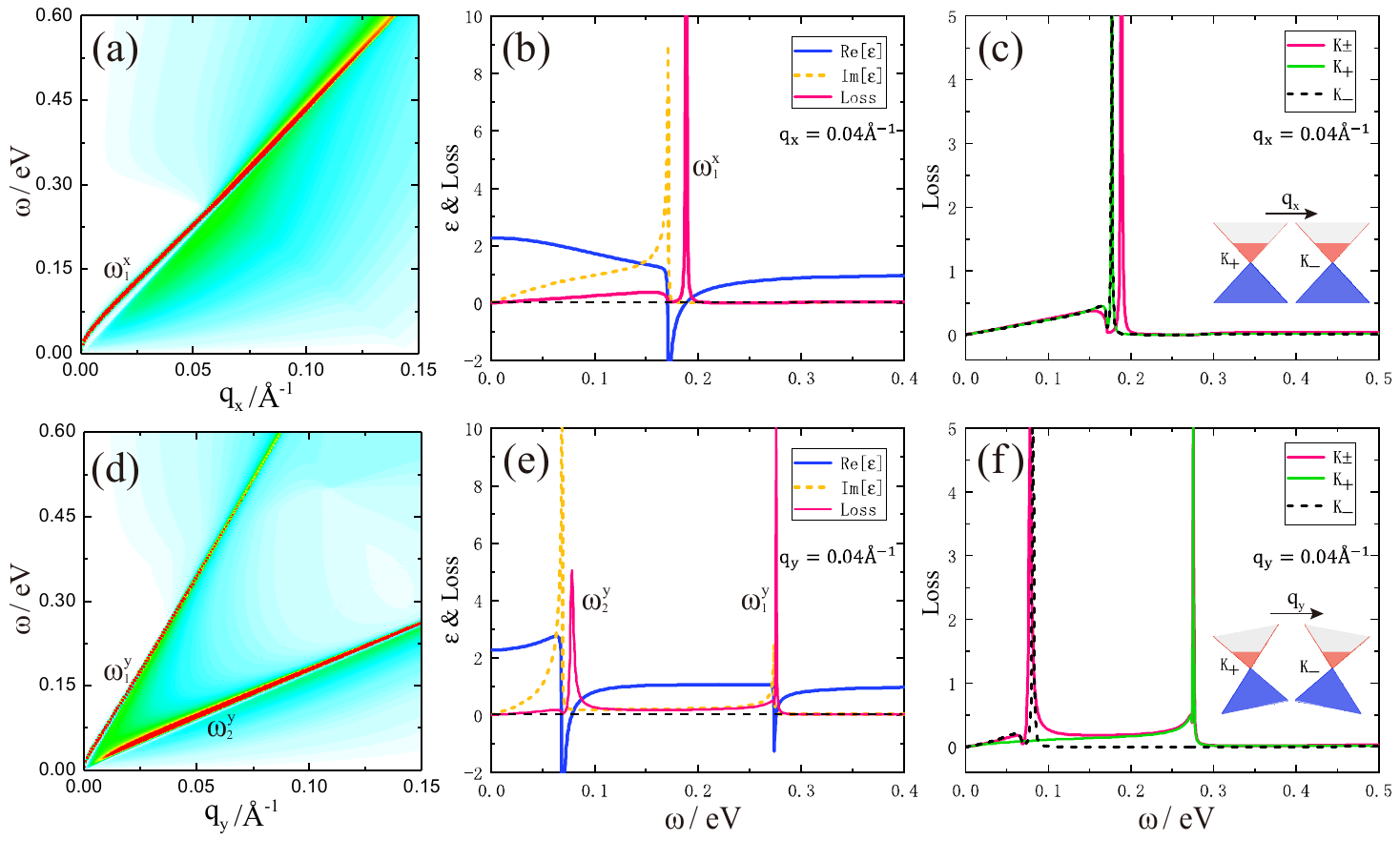}
				\caption{Origin of plasmon for 2D type-I Dirac
					semimetal in terms of the hybridization between the $\mathrm{K}_{+}$ and $\mathrm{K}_{-}$ valleys.
					The wave vector $\boldsymbol{q}$ is (a-c) perpendicular to the tilting
					direction, or (d-f) parallel to the tilting direction. The main parameters
					are taken to be the same as those in main text.} \label{fig:type-IOrigin}
			\end{figure}

			\section{S5. Approximate dispersions for three plasmon modes in the long-wavelength limit}
			
			In this section, we derive the approximate analytical expressions of three plasmon modes in the long-wavelength limit.
			
			\subsection{A. Approximate plasmon dispersions of $\omega_{1}^{x}(q_{x})$ and $\omega_{1}^{y}(q_{y})$}
			
			In this subsection, we derive the approximate plasmon dispersions of conventional gapless  $\sqrt{q}$
			plasmons $\omega_{1}^{x}(q_x)$ and $\omega_{1}^{y}(q_y)$ in the long wavelength limit. In this limit, the intraband overlap becomes unity while the interband overlap vanishes. Consequently, the polarization function for band $\lambda$ around Dirac node $\mathrm{K}_{\chi}$ reduces to 
			\begin{align}
				\Pi_{\lambda}^{\chi}(\boldsymbol{q},\omega)&= \int\frac{d^{2}\boldsymbol{k}}{(2\pi)^{2}}\frac{\delta(E^{\chi}_{\lambda}(\boldsymbol{k})-\mu)}{\omega-\boldsymbol{q}\cdot\nabla E^{\chi}_{\lambda}(\boldsymbol{k})+i\eta}\boldsymbol{q}\cdot\nabla E^{\chi}_{\lambda}(\boldsymbol{k})\notag\\
				&=\int\frac{d^{2}\boldsymbol{k}}{(2\pi)^{2}}\left(1+\frac{\boldsymbol{q}\cdot\nabla E^{\chi}_{\lambda}(\boldsymbol{k})}{\omega}+\cdots\right)\frac{\boldsymbol{q}\cdot\nabla E^{\chi}_{\lambda}(\boldsymbol{k})}{\omega}\delta(E^{\chi}_{\lambda}(\boldsymbol{k})-\mu)
			\end{align} 
			where we retain terms up to $\mathcal{O}(q^2)$. Summing over all band and node indices and performing straightforward algebra to identify the zeros of the dielectric function, we obtain,
			\[
			\begin{aligned}\omega_{1}^{x}(q_{x}) & \approx\text{2\ensuremath{\sqrt{\frac{e^{2}}{\kappa}}}(\ensuremath{
						\pi n_{e})^{1/4}\sqrt{F_{o}^{x}(t)\hbar\upsilon_{F}q_{x}}}},\\
				\omega_{1}^{y}(q_{y}) & \approx2\sqrt{\frac{e^{2}}{\kappa}}(
				\pi n_{e})^{1/4}\sqrt{\text{\ensuremath{F_{o}^{y}(t)}}\hbar\upsilon_{F}q_{y}},
			\end{aligned}
			\]
			where 
			$n_{e}$ denotes the
			carrier density in a single Dirac node as $n_{e}=
			\mu^{2}/(4\pi\hbar^{2}\upsilon_{F}^{2})$.
			Besides, the auxiliary function $F_{o}^{x/y}(t)$ is defined as
			
			\[
			\begin{aligned}F_{o}^{x}(t)= & \frac{1}{\pi}\left[\frac{1}{t^{2}}\frac{1}{\sqrt{t^{2}-1}}\ln\left|\frac{(u_{1}+1)(u_{2}-1)}{(u_{1}-1)(u_{2}+1)}\right|-\frac{\theta_{1}+\theta_{2}}{t^{2}}+\frac{1}{t}\left(\frac{\mathrm{sin}(\theta_{1})}{t\mathrm{cos}(\theta_{1})+1}+\frac{\mathrm{sin}(\theta_{2})}{t\mathrm{cos}(\theta_{2})-1}\right)\right],\\
				F_{o}^{y}(t)= & \frac{1}{\pi}\left[\frac{\sqrt{t^{2}-1}}{t^{2}}\ln\left|\frac{(u_{1}+1)(u_{2}-1)}{(u_{1}-1)(u_{2}+1)}\right|+\frac{\theta_{1}+\theta_{2}}{t^{2}}+\frac{t^{2}-1}{t}\left(\frac{\mathrm{sin}(\theta_{1})}{t\mathrm{cos}(\theta_{1})+1}+\frac{\mathrm{sin}(\theta_{2})}{t\mathrm{cos}(\theta_{2})-1}\right)\right],
			\end{aligned}
			\]
			where
			\[
			\begin{gathered}u_{1}\equiv\sqrt{\frac{t-1}{t+1}}\mathrm{tan}(\frac{\theta_{1}}{2}),\:u_{2}\equiv\sqrt{\frac{t+1}{t-1}}\mathrm{tan}(\frac{\theta_{2}}{2})\\
				\mathrm{cos}(\theta_{1})\equiv\frac{\mu-\hbar\upsilon_{F}\Lambda}{\hbar\upsilon_{F}\Lambda t},\:\mathrm{cos}(\theta_{2})\equiv\frac{\mu+\hbar\upsilon_{F}\Lambda}{\hbar\upsilon_{F}\Lambda t}
			\end{gathered}
			\]
			with $\Lambda$ being the cutoff of wave vector that is always greater
			than $\mu/(t-1)\hbar\upsilon_{F}$.
			
			The approximate solutions of $\omega_{1}^{x}(q_x)$ and $\omega_{1}^{y}(q_y)$
			are compared with the numerical solutions in Figs. 2(a,b) in the main text. It is evident that the numerical solutions of plasmons can be well
			captured by these approximate solutions in the regions of interest.
			
			\subsection{B. Approximate plasmon dispersions of $\omega_{2}^{y}(q_y)$ and $\omega_{3}^{y}(q_y)$}
			
			In this subsection, we derive the approximate analytical expressions of AAPs $\omega_{2}^{y}(q_y)$ and $\omega_{3}^{y}(q_y)$ along the $q_y$ direction in the long-wavelength limit. We restrict to the valley with $\chi=+1$, which contributes dominantly. The energy dispersion is denoted as $E_{\lambda}(\boldsymbol{k})$ for simplicity. The following approximations are employed in calculating the dielectric function:
			\begin{enumerate}
				\item It is assumed that $q_y \ll k_x, k_y$, which allows the long-wavelength expansion
				\begin{equation}
					\sqrt{k_x^2 + (k_y + q_y)^2} \approx k + \frac{k_y q_y}{k},
				\end{equation}
				with $k = \sqrt{k_x^2 + k_y^2}$. In the long-wavelength limit, the dominant contribution is the transition from the narrow hyperbolic strip between two boundaries determined by $E_{\lambda}(\boldsymbol{k}) = \mu$ and $E_{\lambda}(\boldsymbol{k}+\boldsymbol{q})= \mu$ in Fig.\ref{fig:overlap}. 
				
				\item In this narrow hyperbolic strip, the integrand in Eq.(4) of main text can be evaluated in the near-Fermi-surface approximation, where the energy $E_{\lambda}\left(\boldsymbol{k}\right)$ in the integrand can be substituted by the chemical potential, namely, $E_{\lambda}\left(\boldsymbol{k}\right) = \mu$. Additionally, the overlap factor $\mathcal{F}_{\lambda\lambda^{\prime}}^{\chi}\left(\boldsymbol{k},\boldsymbol{k}^{\prime}\right)$ in Eq.(4) of main text is assumed to be a constant $\mathcal{F}$ which is approximately 1 in almost the whole narrow hyperbolic strip.

				\item In the transition region of narrow hyperbolic strip, the integrand can be considered to be independent of $k_x$. After taking both part into account, and the distance between the boundary hyperbolas in the $k_x$ direction is approximately $q_y \sqrt{t^2-1}$ for a large $|\boldsymbol{k}|$. After taking both aspects into consideration, we simplify the integration over $k_x$ as $\int f\left(k_y\right) dk_x \approx 2 q_y \sqrt{t^2-1}f(k_y) $.
				
			\end{enumerate}

			\begin{figure}[htbp]
				\includegraphics[scale=0.6]{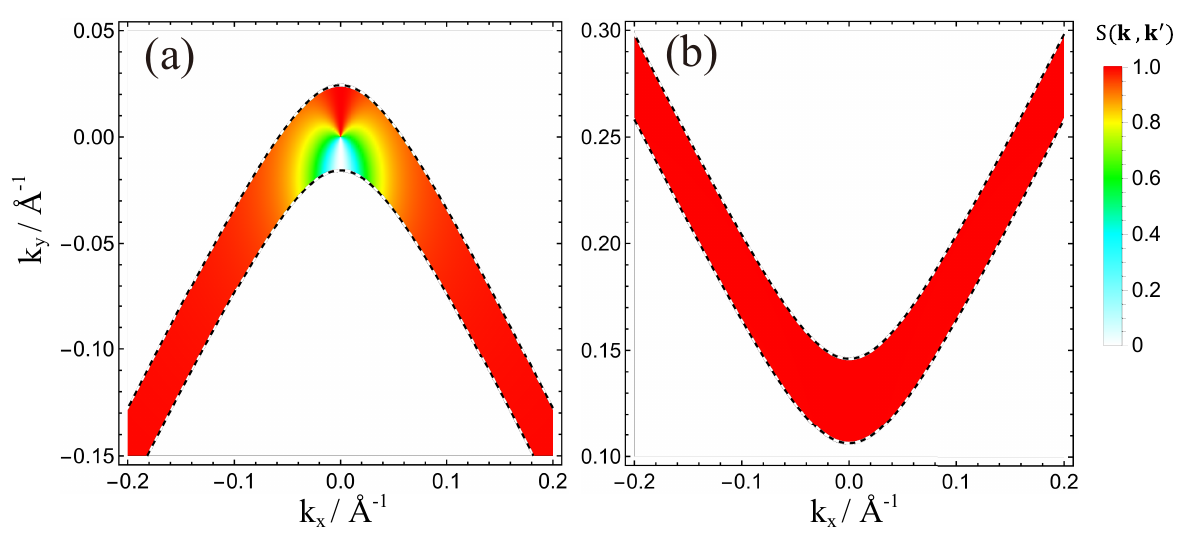}
				\caption{The intraband correlations demonstrated by $\mathrm{S}(\boldsymbol{k},\boldsymbol{k}^{\prime})=\mathcal{F}_{\lambda\lambda^{\prime}}^{\chi}\ensuremath{\Theta\left[E_{\lambda}^{\chi}(\boldsymbol{k}^{\prime})-\mu\right]}\ensuremath{\Theta\left[\mu-E_{\lambda^{\prime}}^{\chi}(\boldsymbol{k})\right]}$ for $\chi=+$ and $\lambda^{\prime}=\lambda$ in the limit $q_y \ll k_x, k_y$. The pocket index is set to be $\lambda=+$ for (a) and $\lambda=-$ for (b). 
				}
				\label{fig:overlap}
			\end{figure}

			\subsubsection{1. Energy Difference}

			From the energy dispersion of Type-II Dirac fermion $E_{\lambda}(\boldsymbol{k}) = v_t k_y +\lambda v_F \sqrt{k_x^2 + k_y^2}$, the energy difference for a transition in the $q_y$-direction is:
			\begin{equation}
				\Delta E = E_{\lambda}(\boldsymbol{k}+\boldsymbol{q}) - E_{\lambda}(\boldsymbol{k}) \approx 
				v_t q_y +\lambda v_F \frac{k_y q_y}{k},
			\end{equation}
			to the leading order of $q_y$.
			
			Using the near-Fermi-surface approximation $E_{\lambda}(\boldsymbol{k}) = \mu$, the relation for $k$ is obtained as:
			\begin{equation}
				k = \frac{\mu - v_tk_y}{\lambda v_F}.
			\end{equation}
			And then
			\begin{equation}
				\Delta E\approx v_t q_y +\lambda v_F \frac{k_y q_y}{k}= v_t q_y + \lambda v_F \frac{k_y q_y}{\frac{\mu - v_t k_y}{\lambda v_F}}
				= v_t q_y + \lambda v_F^2 \frac{k_y q_y}{\mu - v_t k_y}.
			\end{equation}

			\subsubsection{2. Polarization Function Calculation}
			The integration for the polarization function $\Pi_{+}^{\lambda}(q_y, \omega)$ involves a $k_y$-dependent denominator. The integrations for the P pocket ($\lambda=+$) and the N pocket ($\lambda=-$) are given by:
			\begin{equation}
				\Pi_{+}^{+}(q_y, \omega)\approx \frac{2 q_y\mathcal{F}^{\prime} }{(2 \pi)^2} \int_{-\Lambda}^{\frac{\mu}{v_t+v_F}} \frac{1}{\omega - v_t q_y - \frac{v_F^2 k_y q_y}{\mu - v_t k_y}} d k_y,
			\end{equation}
			\begin{equation}
				\Pi_{-}^{+}(q_y, \omega) \approx \frac{2 q_y\mathcal{F}^{\prime}}{(2 \pi)^2} \int_{\frac{\mu}{v_t-v_F}}^{\Lambda} \frac{1}{\omega - v_t q_y - \frac{v_F^2 k_y q_y}{\mu - v_t k_y}} d k_y,
			\end{equation}
			where $\mathcal{F}^{\prime}=\mathcal{F} \sqrt{t^2-1}$. For the P pocket, the upper limit of integration in $\Pi_{+}^{+}(q_y, \omega)$ is approxiately 0, namely $\frac{\mu}{v_t+v_F}\approx0$, by considering that it is much smaller than $\Lambda$. By contrast, for the N pocket, the lower limit $\frac{\mu}{v_t-v_F}$ could be large and hence  non-negligible. The result of this integration includes logarithmic terms of the form $\log[\dots]$, abbreviated as ``$\mathcal{L}$''. For simplicity, we treated it as a constant with a value typically ranging from -5 to 5 when solving the dispersion equation. 
			
			The plasmon dispersion can be obtained by finding the zeros of the dielectric equation,
			\begin{equation}
				1 - \frac{2 \pi e^{2}}{\kappa q_y} \left\{\frac{\mathcal{F}^{\prime} q_y  \left[v_F^2 q_y \mu \mathcal{L} +  \left((v_F^2 - v_t^2)q_y + v_t \omega\right)v_t\Lambda \right]}{2 \pi^2 \left((v_F^2 - v_t^2)q_y + v_t \omega\right)^2} \right\}  = 0.
			\end{equation}
			
			\subsubsection{3. Dispersion Relations}
			By solving the dielectric equation, we obtain the dispersion relations for the two modes:
			
			\begin{equation}
				\omega_2^y(q_y) = \frac{e^{2}\mathcal{F}^{\prime} v_t^2\Lambda  +2 \pi \kappa v_t (v_t^2-v_F^2)q_y -\sqrt{\left(e^{2}\mathcal{F}^{\prime} v_t^2\Lambda\right)^2+ 4 \pi\kappa v_F^2 v_t^2 \mathcal{F}^{\prime} \mathcal{L} \mu q_y }}{2 \pi \kappa v_t^2 },
			\end{equation}
			
			\begin{equation}
				\omega_3^y(q_y) = \frac{e^{2}\mathcal{F}^{\prime} v_t^2\Lambda  +2 \pi \kappa v_t (v_t^2-v_F^2)q_y +\sqrt{\left(e^{2}\mathcal{F}^{\prime} v_t^2\Lambda\right)^2+ 4 \pi\kappa v_F^2 v_t^2 \mathcal{F}^{\prime} \mathcal{L} \mu q_y }}{2 \pi \kappa v_t^2   }.
			\end{equation}
			
			\begin{enumerate}
				\item The gapless mode $\omega_2^{y}(q_y)$ takes the following form
				\begin{equation}
					\omega_{2}^{y}(q_y) \approx \left[t-\frac{1}{t}\left(1+\frac{\mathcal{L}}{t}\frac{|\mu|}{v_{F}\Lambda}\right)\right]v_{F}q_y,
				\end{equation}
				which is characterized by a linear dispersion relation, and its frequency is independent of the effective dielectric constant $\kappa$ of the background. 
				
				\item The gapped mode $\omega_3^{y}(q_y)$  takes the analytical form
				\begin{equation}
					\omega_{3}^{y}(q_y) \approx\sqrt{t^2-1}\frac{e^2\Lambda}{\pi\kappa}+ \left[t-\frac{1}{t}\left(1-\frac{\mathcal{L}}{t}\frac{|\mu|}{v_{F}\Lambda}\right)\right]v_{F}q_y,
				\end{equation}
				which exhibits a finite energy gap at $q_y \to 0$. This gap is determined by the tilting paramter $t$, the cutoff of wave vector $\Lambda$, the overlap factor $\mathcal{F}\sim 1$, and the dielectric constant $\kappa$.
				
			\end{enumerate}

			Fitting with the numerical results, we set $\mathcal{L}=1/2$. Consequently, the two AAP modes take 
			\begin{align}
				\omega_{2}^{y}(q_y) &\approx \left[t-\frac{1}{t}\left(1+\frac{|\mu|}{2tv_{F}\Lambda}\right)\right]v_{F}q_y,\\
				\omega_{3}^{y}(q_y) &\approx \sqrt{t^2-1}\frac{e^2\Lambda}{\pi\kappa}+\left[t-\frac{1}{t}\left(1-\frac{|\mu|}{2tv_{F}\Lambda}\right)\right]v_{F}q_y.
			\end{align}

			\subsection{C. Approximate plasmon dispersions of $\omega_{2}^{x}(q_x)$ and $\omega_{3}^{x}(q_x)$}
			
			When the wave vector is along $q_x$-direction, the dispersion equation is similarly solved for $\varepsilon(q_x, \omega) = 0$. Consequently, the two AAP modes take 
			\begin{align}
				\omega_{2}^{x}(q_x) &\approx \sqrt{1-\frac{1}{t^2}\left(1+\frac{|\mu|}{t v_{F}\Lambda}\right)^2}v_{F}q_x,\\
				\omega_{3}^{x}(q_x) &\approx \frac{e^2\Lambda}{2\pi\kappa}+\sqrt{1-\frac{1}{t^2}\left(1-\frac{|\mu|}{t v_{F}\Lambda}\right)^2}v_{F}q_x.
			\end{align}

			\begin{table}[th]
				\begin{tabular}{|l|lll|lll|}
					\hline
					& \multicolumn{3}{c|}{Plasmons}                                                                  & \multicolumn{3}{c|}{Origin}                                                                                                                                                                                          \\ \hline
					\multicolumn{1}{|c|}{\multirow{3}{*}{3D}} & \multicolumn{3}{l|}{Gapped plasmon initiated from $(0,\omega_{c}^{y})$.}                      & \multicolumn{3}{l|}{Regular mode in 3D system from intraband correlation.}                                                                                                                                           \\ \cline{2-7} 
					\multicolumn{1}{|c|}{}                    & \multicolumn{3}{l|}{Damped plasmon.}                                                          & \multicolumn{3}{l|}{Out-of-phase oscillation of the electron fluids in different nodes.}                                                                                                                             \\ \cline{2-7} 
					\multicolumn{1}{|c|}{}                    & \multicolumn{3}{l|}{Gapless plasmon initiated from $(q_{c}^{y},0)$.}                          & \multicolumn{3}{l|}{\begin{tabular}[c]{@{}l@{}}Novel mode from out-of-phase oscillations of the intranode \\electron-hole pockets.\end{tabular}}                                                                                                              \\ \hline
					\multirow{6}{*}{2D}                       & \multicolumn{3}{l|}{\begin{tabular}[c]{@{}l@{}}$\omega_{1}$: gapless plasmon initiated from origin\\ in the $q-\omega$ plane. \end{tabular}}                     & \multicolumn{3}{l|}{Regular mode in 2D system from intraband correlation.}                                                                                                                                           \\ \cline{2-7} 
					& \multicolumn{3}{l|}{$\omega_{2}$: acoustic plasmon.}                                           & \multicolumn{3}{l|}{\begin{tabular}[c]{@{}l@{}}Novel mode from hybridization of two pockets $\mathrm{K}_+^\mathrm{P}$ and $\mathrm{K}_+^\mathrm{N}$ on $\mathrm{K}_+$ \\ valley followed by the modification of $\mathrm{K}_-$ valley. In contrast, no \\plasmon on the isolated $\mathrm{K}_-$ valley.\end{tabular}} \\ \cline{2-7} 
					& \multicolumn{3}{l|}{$\omega_{3}$: gapped plasmon initiated from $(q_{c}^{y},\omega_{c}^{y})$.} & \multicolumn{3}{l|}{\begin{tabular}[c]{@{}l@{}}Novel mode living deep into the intraband SPE region due to the \\ enhancement by open Fermi surface.\end{tabular}}                                 \\ \hline
				\end{tabular}
				\caption{Plasmons along the tilting direction in 2D and 3D Type-II  Dirac cones.}\label{SM_TAB1}
			\end{table}
			
			\begin{table}[th]
				\begin{tabular}{|l|lll|lll|}
					\hline
					& \multicolumn{3}{c|}{Plasmons}                                                                  & \multicolumn{3}{c|}{Origin}                                                                                                                                                                                                                                                                    \\ \hline
					\multicolumn{1}{|c|}{3D} & \multicolumn{3}{l|}{Gapped plasmon initiated from $(0,\omega_{c}^{x})$.}                      & \multicolumn{3}{l|}{Regular mode in 3D system from intraband correlation.}                                                                                                                                                                                                                     \\ \hline
					\multirow{6}{*}{2D}      & \multicolumn{3}{l|}{\begin{tabular}[c]{@{}l@{}}$\omega_{1}$: gapless plasmon initiated from origin \\in the $q-\omega$ plane.\end{tabular}}                     & \multicolumn{3}{l|}{Regular mode in 2D system.}                                                                                                                                                                                                                                                 \\ \cline{2-7} 
					& \multicolumn{3}{l|}{$\omega_{2}$: acoustic plasmon.}                                           & \multicolumn{3}{l|}{\begin{tabular}[c]{@{}l@{}}Novel mode from hybridization of two pockets $\mathrm{K}_+^\mathrm{P}$ and $\mathrm{K}_{+}^\mathrm{N}$ on $\mathrm{K}_{+}$\\ valley followed by the modification of $\mathrm{K}_{-}$ valley. Notably, plamson \\is still significant on the isolate $\mathrm{K}_{-}$ valley, and mechanism of\\ hybridization works well on it.\end{tabular}} \\ \cline{2-7} 
					& \multicolumn{3}{l|}{$\omega_{3}$: gapped plasmon initiated from $(q_{c}^{x},\omega_{c}^{x})$.} & \multicolumn{3}{l|}{\begin{tabular}[c]{@{}l@{}}Novel mode living deep into the intraband SPE region due to the \\enhancement by open Fermi surface.\end{tabular}}                                                                                                         \\ \hline
				\end{tabular}
				\caption{Plasmons perpendicular to the tilting direction in 2D and 3D Type-II  Dirac cones.}\label{SM_TAB2}
			\end{table}
			
			\section{S6. Comparison between plasmons in 2D and 3D type-II Dirac cones} 
			
			We dealt with the two-dimensional over-tilted Dirac cone instead of the three-dimensional counterpart \cite{Agarwal2020}. Multiple plasmon excitations are found in both the 2D and 3D type-II Dirac semimetals with multiple Fermi pockets. However, due to the different dimensionalities and Coulomb screening, some plasmon excitations have distinct features. For concreteness, we make a comparison between plasmons in 2D and 3D type-II Dirac cones in Table. \ref{SM_TAB1} and \ref{SM_TAB2}.

\end{widetext}


\begin{thebibliography}{99}
\bibitem {MuPlasmon} For a finite Fermi level apart from the Dirac point, the qualitative features of three plasmons due to over-tilting of Dirac cones are not sensitive to the magnitude of the Fermi level. When the Fermi level crosses the Dirac point, the lowest frequency plasmon almost disappears due to over damping \cite{SMs}.
\bibitem {TBPlasmon} The calculation based on the equivalent tight-binding model also yields three plasmon modes and is in line with the results in Fig. (2) from the effective $k\cdot p$ model \cite{SMs}.
\end{thebibliography}
\end{document}